\documentclass[graybox]{svmult}

\usepackage{mathptmx}       
\usepackage{helvet}         
\usepackage{courier}        
\usepackage{type1cm}        
\usepackage{makeidx}         
\usepackage{graphicx}        
\usepackage{multicol}        
\usepackage[bottom]{footmisc}
\usepackage{amsmath}
\usepackage{amssymb}
\usepackage{bm}
\usepackage{esint}

\makeindex             

\begin{document}

\title*{Entropic Uncertainty Relations\\in Quantum Physics}
\author{Iwo Bialynicki-Birula and {\L}ukasz Rudnicki}
\institute{Center for Theoretical Physics, Polish Academy of Sciences, Al. Lotnik{\'o}w 32/46\\
02-668 Warszawa, Poland\\
\email{birula@cft.edu.pl, rudnicki@cft.edu.pl}}

%
%
\maketitle

\abstract{\;Uncertainty relations have become the trademark of quantum theory since they were formulated by Bohr and Heisenberg. This review covers various generalizations and extensions of the uncertainty relations in quantum theory that involve the R\'enyi and the Shannon entropies. The advantages of these entropic uncertainty relations are pointed out and their more direct connection to the observed phenomena is emphasized. Several remaining open problems are mentioned.}

\section*{Introduction}

In recent years we have seen many applications of the R\'enyi and Shannon entropies in many fields from biology, medicine, genetics, linguistics, and economics to electrical engineering, computer science, geophysics, chemistry, and physics. In particular, the R\'enyi entropy has been widely used in the study of quantum systems. It was used in the analysis of quantum entanglement \cite{asi,bez,bcehas,gl,ter}, quantum communication protocols \cite{gl1,rgk}, quantum correlations \cite{lnp}, quantum measurement \cite{bg}, and decoherence \cite{kh}, multiparticle production in high-energy collisions \cite{bc,bcz,bcz1}, quantum statistical mechanics \cite{mw}, pattern formation \cite{cmbh,cbh}, localization properties of Rydberg states \cite{arb} and spin systems \cite{gz,vc}, in the study of the quantum-classical correspondence \cite{dmfs}, in electromagnetic beams \cite{beams}, and the localization in phase space \cite{salcedo,vp}.

Our aim in this review is to use the Shannon and R\'enyi entropies to describe the limitations on the available information that characterizes the states of quantum systems. These limitations in the form of mathematical inequalities have the physical interpretation of the uncertainty relations. We will not enter here into the discussion (cf. \cite{bz,cgt}) of a fundamental problem: which (if any) entropic measure of uncertainty is really most adequate in the analysis of quantum mechanical measurements. The uncertainty relations discussed in this paper are valid as mathematical inequalities, regardless of their physical interpretation. Since this is a review, we felt free to use the results of our previous investigations. Of course, such self-plagiarism would be highly unethical in an original publication.

\section{Information entropy as a measure of uncertainty}

Statistical complexity, a tool presented and explored in this book is based on information entropy. This concept was introduced by Claude Shannon \cite{shan} in 1948 and grew up to have many applications. There are also several generalizations and extensions which we will discuss in our contribution. We shall focus on the uncertainty, a notion closely connected with the information entropy but a little bit wider and having different meanings depending on a context. We shall use here the term uncertainty as a measure of missing information. The use of the information entropy as a measure of uncertainty becomes then very natural. All we have to do is to reverse the sign. The lack of information --- negative information --- is the uncertainty. In simple terms, we can measure the capacity of an empty tank by measuring the volume of water filling this tank. Therefore, the uncertainty or missing information can be measured in exactly the same manner as the information is measured. We shall exploit this point of view showing that the measure of uncertainty based on the information entropy may be used to replace famous quantum mechanical uncertainty relations. Moreover, we shall argue now that this method of expressing the uncertainty relations is much closer to the spirit of quantum theory.

\subsection{Standard deviation}

In classical physics the positions and momenta of every particles can be determined without any fundamental limitations. In that case, when we talk about an uncertainty of position or momentum of a particle, we mean an uncertainty caused by a lack of precision of our measuring instruments. When we measure some observable $Q$ we usually repeat this measurement many times obtaining a distribution of values. We apply then various standard procedures, well known to experimentalists, to extract the ``best value'' of the measured quantity. A crude, but often a sufficient procedure is to evaluate an average value $\langle Q\rangle $ of $Q$. We can also calculate the standard deviation:
\begin{align}
\sigma_{Q}=\sqrt{\langle \left(Q-\langle Q\rangle \right)^{2}\rangle },\label{Q1}
\end{align}
which we treat as a measure of uncertainty of $\langle Q\rangle $. It is the uncertainty connected directly with a given experimental setup and has no fundamental significance. A more skilled experimentalist will be able to reduce this uncertainty.

In quantum physics we face a dramatic change because we have to deal with a spread of measured values which is of fundamental nature. The uncertainty in most measurements (there are some exceptions) cannot be indefinitely decreased. This is due, at least at this stage of our theoretical understanding, to the probabilistic nature of the quantum world. For example, the state of a single nonrelativistic quantum particle can be described by a wave function $\psi\left({\bm r}\right)$. The square of the modulus of this wave function $\rho\left({\bm r}\right)=|\psi\left({\bm r}\right)|^{2}$ determines the probability distribution of the position of the particle. Since the probability to find the particle anywhere must be equal to 1, the function $\psi\left({\bm r}\right)$ must be normalized according to:
\begin{align}\label{norm}
\int_{\mathbb{R}^3}d^3r\,|\psi\left({\bm r}\right)|^2 = 1.
\end{align}
This probabilistic distribution of values at our disposal is not connected with some faults of our measuring procedure but it is an intrinsic property --- the spread of values of the position ${\bm r}$ cannot be avoided. The classical position ${\bm r}_{cl}$ can at best be associated with the average value
\begin{align}
{\bm r}_{cl}=\langle {\bm r}\rangle =\int_{\mathbb{R}^3}d^3r\,{\bm r}\,\rho\left({\bm r}\right).\label{Q2}
\end{align}
Having a probability distribution $\rho\left({\bm r}\right)$ we can proceed according to the rules of statistics and calculate the standard deviation, say $\sigma_{x}$, that characterizes the spread of the values of the coordinate $x$,
\begin{align}\label{dispx}
\sigma_{x}=\left[\int_{\mathbb{R}^3}d^3r\,(x-\langle x\rangle)^2\,\rho\left({\bm r}\right)\right]^{1/2}.
\end{align}

After the Fourier transformation, we can obtain another description of {\em the same state}
\begin{align}
\tilde{\psi}\left({\bm p}\right)=\int_{\mathbb{R}^3}\frac{d^3r}{(2\pi\hbar)^{3/2}}\,
e^{-i{\bm p}\cdot{\bm r}/\hbar}\psi\left({\bm r}\right).\label{Q3}
\end{align}
The Fourier transform $\tilde{\psi}\left({\bm p}\right)$ of a wave function, according to the rules of quantum mechanics, gives the probability distribution in momentum space $\tilde{\rho}\left({\bm p}\right)=|\tilde{\psi}\left({\bm p}\right)|^{2}$. Note, that $\tilde{\rho}\left({\bm p}\right)$ is not the Fourier transform of $\rho({\bm r})$. Due to the Plancherel theorem for the Fourier transform this probability distribution is normalized since the original function was normalized as expressed by Eq.~(\ref{norm}). Using this probability distribution we can calculate $\sigma_{p_x}$ the standard deviation of the $p_x$ component of the momentum,
\begin{align}\label{dispp}
\sigma_{p_x}=\left[\int_{\mathbb{R}^3}d^3p\,(p_x-\langle p_x\rangle)^2\,{\tilde\rho}\left({\bm p}\right)\right]^{1/2}.
\end{align}

Even though for different states both standard deviations $\sigma_{x}$ and $\sigma_{p_x}$ can be arbitrarily small when treated separately, they become correlated when calculated for the same state. This correlation is usually expressed by the Heisenberg uncertainty relation,
\begin{align}\label{Q4}
\sigma_{x}\sigma_{p_x}\geq\frac{\hbar}{2}.
\end{align}
The bound in the inequality ({\ref{Q4}) is saturated by Gaussian wave functions. Heisenberg explained this relation in the following words \cite{heis}:

``the more accurately the position is known, the less accurately is the momentum determined and {\em vice versa}''.

Quantum mechanics gave an important reason to study the inherent incompleteness (or uncertainty) of our information about quantum objects. However, in our opinion, the expression of the uncertainty in terms of standard deviations is too much ingrained in classical physics to be of importance at a more fundamental level. We shall argue in the next section that the measure of uncertainty in terms of information entropy is much more appropriate in the quantum world. We shall prove the uncertainty relations in quantum mechanics expressed in terms of the information entropy. Finally, in the last section we introduce further generalizations of these ideas. We begin with a critical analysis of the standard deviations as measures of uncertainty.

\subsection{Limitations on the use of standard deviations}\label{E1}

As we have already noticed, the standard deviation is a well known and widely accepted measure of uncertainty. However it is rarely mentioned that it is not an ideal tool, failing even in very simple situations. The reason, why we should be careful while using the standard deviation, is explained with the help of the following examples \cite{ibbren}.

\subsubsection{Example I
}
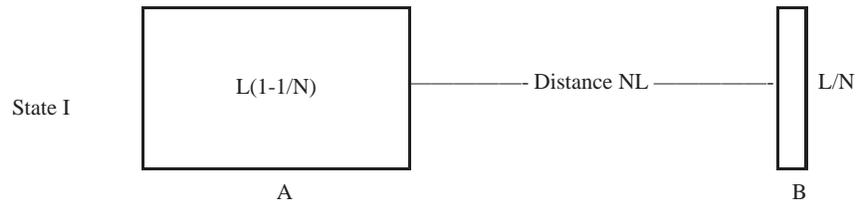
\begin{figure}
\hspace{2cm}
\begin{picture}(340,100)(600,-130)
\put(550,-80){State I} \thicklines \put (600,-100){\framebox(100,60){L(1-1/N)}} \put (700,-70){---------------- Distance NL ----------------} \put(840,-100){\framebox(10,60){}}
\put (855,-70){L/N}
\put (650,-112){A}
\put (845,-112){B}
\end{picture}
\caption{The particle is to be found mostly in a central region. In addition, there is a tiny probability that the particle can be found far away.}\label{fig1}
\end{figure}

Let us specify two regions on the real axis (see Fig.~\ref{fig1}). The first region is the line segment $A=\left[L\left(N+1/N\right),L\left(N+1\right)\right]$ with its length equal to $L\left(1-1/N\right)$ where $L$ is the unit length and $N$ is a large number. The second region B is the line segment with its length equal to $L/N$ separated from the first region by the large distance $NL$. In the next step let us assume that we have the probability distribution of the form (the boxes in Fig.~\ref{fig1} represent the probability distribution):
\begin{align}
\rho(x)=\begin{cases} 1/L & \qquad x\in A\\ 1/L & \qquad x\in B\\ 0 & \quad\;\textrm{elsewhere}\end{cases}\label{Q5}
\end{align}
The standard deviation of the variable $x$ calculated in this case is:
\begin{align} \sigma_{x}^{2}\left(\textrm{I}\right)=L^{2}\left(N-\frac{1}{N}+\frac{1}{12}\right),\label{Q6}
\end{align}
and for sufficiently large $N$ we obtain:
\begin{align}
\sigma_{x}\left(\textrm{I}\right)\approx L\sqrt{N}.\label{Q7}
\end{align}
Therefore, $\sigma_{x}$ tends to infinity with $N\to\infty$, while common sense predicts that the uncertainty should remain finite. So, why the uncertainty measured by the standard deviation grows with $N$? It is simply the effect of a growing distance between the two regions. Standard deviation is based on the second moment which is very sensitive to the values of an observable lying far away from the average. This is the first flaw of the standard deviation. Another one appears in the second example \cite{ibbren}.

\subsubsection{Example II}\label{E3}

\begin{figure}
\hspace{1.0cm}

\begin{picture}(240,40)(600,-100)
\put(600,-80){State IIA} \thicklines \put (650,-90){\framebox(60,30){I}} \put (710,-90){\framebox(60,30){II}} \put (770,-90){\framebox(60,30){III}} \put (830,-90){\framebox(60,30){IV}}
\end{picture}
\caption{The particle is localized with a uniformly distributed probability in a box of length L.}

\begin{picture}(240,80)(600,-130)
\put(600,-94){State IIB} \thicklines \put (650,-120){\framebox(60,30){I}} \put (650,-90){\framebox(60,30){II}} \put (830,-120){\framebox(60,30){III}} \put (830,-90){\framebox(60,30){IV}}
\end{picture}
\caption{The particle is localized with equal probabilities in two smaller boxes each of length $L/4$.}
\end{figure}
In the previous example the size of the space where the probability distribution is constant is preserved. Now we omit this condition to present, one more time, inadequateness of the standard deviation and prepare the reader for the solution of this problem. Let us assume two different cases (see Figs~2 and 3). In the case A we take:
\begin{align}
\rho_{A}(x)=\begin{cases} 1/L & \qquad x\in\left[0,L\right]\\ 0 & \qquad\textrm{elsewhere}\end{cases},\label{Q8}
\end{align}
in the case B we have:
\begin{align}
\rho_{B}(x)=\begin{cases} 2/L & \qquad x\in\left[0,L/4\right]\\ 2/L & \qquad x\in\left[3L/4,L\right]\\ 0 & \qquad\textrm{elsewhere}\end{cases}.\label{Q9}
\end{align}
It is easy to see that in the case A the size of the space where the particle can be found is $L$, while in the case B this size is $L/2$. In the case B we know more about position than in the case A. According to this obvious argument it is rather disappointing that the uncertainty of the position is greater in the case B,
\begin{subequations}
\begin{align}
\sigma_{x}\left(\textrm{IIA}\right)&=\frac{L}{\sqrt{12}},\label{Q10}\\ \sigma_{x}\left(\textrm{IIB}\right)&=\sqrt{\frac{7}{4}}\frac{L}{\sqrt{12}}.\label{Q11}
\end{align}
\end{subequations}
This is another manifestation of the fact that the gap between two regions (line segment $\left[L/4,3L/4\right]$) where the particle cannot be found contributes to the standard deviation.

\subsection{Uncertainty of measurements}

In all natural sciences measurements play a fundamental role. Every measurement has some uncertainty, so we want our theoretical tools to be directly correlated with our knowledge about that uncertainty. On the other hand, as it has been aptly stressed by Peres \cite{ap}, ``The uncertainty relation such as $\sigma_{x}\sigma_{p}\geq\hbar/2$ is not a statement about the accuracy of our measuring instruments.'' The Heisenberg uncertainty relations in their standard form (\ref{Q4}) completely ignore these problems. In what follows we show how to incorporate the quantum mechanical restrictions on the information obtained in an experiment. We shall consider, however, only the ideal situation when the restrictions on the information gathered in the process follow from the statistical nature of quantum mechanics. Other sources of information loss will not be taken into account. Still, the real experimental errors will be, to some extent, represented by the assumed precision of the measurements. To start, we shall rephrase our quotation from Heisenberg by using the notion of information:

``the {\em more information} we have about the position, the {\em less information} we can acquire about the momentum and vice versa''.

\subsection{Shannon entropy}

Since we know precisely how to measure information, we may try to give a rigorous mathematical relation that will reflect this statement. Shannon connected the measure of the information content with a probability distribution. The use of the information entropy fits perfectly the statistical nature of quantum mechanical measurements. All we have to do is to insert the set of probabilities obtained from quantum mechanics into the famous Shannon formula for the information entropy. Considering our mathematical applications, we shall use this formula with natural logarithms, not with logarithms to the base 2,
\begin{align}\label{cs}
H = -\sum_k p_k\ln p_k.
\end{align}
We shall now associate information about the experimental errors with the probabilities $p_k$. The proposal to associate the Shannon entropy with the partitioning of the spectrum of a physical observable was made by Partovi \cite{hp}. As an illustration, let us consider the measurements of the position. For simplicity, we shall consider a one dimensional system and assume that the experimental accuracy $\delta x$ is the same over the whole range of measurements. In other words, we divide the whole region of interest into bins --- equal segments of size $\delta x$. The bin size $\delta x$ will be viewed as a measure of the experimental error.

Each state of a particle whose position is being measured can be associated with a histogram generated by a set of probabilities calculated according to the rules of quantum mechanics. For a pure state, described by a wave function $\psi(x)$, the probability to find the particle in the $i$-th bin is
\begin{align}\label{xbin}
q_i=\int_{\left(i-1/2\right)\delta x}^{\left(i+1/2\right)\delta x}\!\!dx\,\rho(x).
\end{align}
The uncertainty in position for the bin size $\delta x$ is, therefore, given by the formula
\begin{align}\label{csx}
H^{(x)} = -\sum_{i=-\infty}^{\infty} q_i\ln q_i.
\end{align}
For the experimental setup characterized by the bin size $\delta x$ the uncertainty of the measurement is the lowest when the particle is localized in just one bin. The probability corresponding to this bin is equal to 1 and the uncertainty is zero. In all other cases, we obtain some positive number --- a measure of uncertainty. When the number of bins is finite, say $N$, the maximal value of the uncertainty is $\ln N$ --- the probability to find the particle in each bin is the same, equal to $1/N$.

The uncertainty in momentum is described by the formula (\ref{cs}) in which we substitute now the probabilities associated with momentum measurements. Since most of our considerations will have a purely mathematical character, from now on we shall replace momentum by the wave vector ${\bm k}={\bm p}/\hbar$. In this way we get rid of the Planck constant in our formulas. In physical terminology this means that we use the system of units in which $\hbar=1$. According to quantum mechanics, the probability to find the momentum $k_x$ in the $j$-th bin is
\begin{align}\label{kbin}
p_j=\int_{\left(j-1/2\right)\delta k}^{\left(j+1/2\right)\delta k}\!\!dk\,{\tilde\rho}(k),
\end{align}
where ${\tilde\rho}(k)$ is the modulus squared of the one-dimensional Fourier transform ${\tilde\psi}(k)$ of the wave function,
\begin{align}\label{ftk}
{\tilde\psi}(k)=\int_{-\infty}^\infty\frac{dx}{\sqrt{2\pi}}e^{-ikx}\psi(x),
\end{align}
and $\delta k$ is the resolution of the instruments measuring the momentum in units of $\hbar$. The uncertainty in momentum measured with the use of the Shannon formula is constructed in the same way as the uncertainty in position,
\begin{align}\label{csp}
H^{(k)} = -\sum_{j=-\infty}^{\infty} p_j\ln p_j.
\end{align}

\subsubsection{Examples}

The uncertainties measured with the use of the formulas (\ref{csx}) and (\ref{csp}) do not suffer from the deficiencies described by our examples I and II. One can easily check that in the case I the uncertainty caused by the smaller region does not contribute in the limit when $N\to\infty$. In turn, in the case II when $\delta x=L/4$ the uncertainty in the state $A$ is equal to $2\ln 2$ while in the state $B$ it is equal to $\ln 2$. Thus, as expected, the position uncertainty in the state $A$ is greater (we have less information) than in the state $B$.

Now, we shall give one more very simple example that clearly shows the merits of entropic definition of uncertainty. Let us consider a particle in one dimension localized on the line segment $\left[-a,a\right]$. Assuming a homogeneous distribution, we obtain the following wave function:
\begin{align}
\psi_a(x)=\begin{cases} 1/\sqrt{2a} & \qquad x\in\left[-a,a\right]\\ 0 & \qquad{\rm elsewhere}\end{cases},\label{Q36}
\end{align}
The Fourier transform of $\psi_a(x)$ is:
\begin{eqnarray}\label{ft}
{\tilde\psi}_a(k)=\sqrt{\frac{1}{\pi a}}\frac{\sin(ak)}{k}.
\end{eqnarray}
This leads to the following probability distribution in momentum space:
\begin{align}
\tilde{\rho}_a(k)=\frac{1}{\pi ak^{2}}\sin^2(ak).\label{Q39}
\end{align}
The standard uncertainty relation (\ref{Q4}) for this state is meaningless because the second moment of $\tilde{\rho}_a(k)$ is infinite. However the uncertainty in momentum measured with the Shannon formula is finite. Taking for simplicity the bin size in momentum as $\delta k=2\pi/a$, we obtain the following expression for the probability to find the momentum in the $j$-th bin:
\begin{align}\label{pj}
p_j&=\frac{1}{\pi a}\int_{(j-1/2)\delta k}^{(j+1/2)\delta k}\!\!dk\,\frac{\sin^2(a k)}{k^2}\nonumber\\
&=\frac{1}{\pi}\int_{2\pi(j-1/2)}^{2\pi(j+1/2)}\!\!d\eta\,\frac{\sin^2(\eta)}{\eta^2}\nonumber\\
&=\frac{1}{\pi}\int_{2\pi(2j-1)}^{2\pi(2j+1)}\!\!d\eta\,\frac{\sin(\eta)}{\eta}\nonumber\\
&=\frac{{\rm Si}[(4j+2)\pi]}{\pi}-\frac{{\rm Si}[(4j-2)\pi]}{\pi},
\end{align}
where {\rm Si} is  the integral sine function. The uncertainty in momentum is obtained by evaluating numerically the sum (\ref{csp}) which gives $H^{(k)}=0.530$. To obtain the position uncertainty, we take just two bins ($\delta x=a$), so that $H^{(x)}=\ln2$. The sum of these two uncertainties is about 1.223.

\subsection{Entropic uncertainty relations}\label{mom}

We have shown that the use of the Shannon formula gives a very sensible measure of uncertainties that takes into account the resolution of measuring devices. The situation becomes even more interesting when we consider measurements on the same quantum state of two ``incompatible'' properties of a particle. The term incompatible has a precise meaning in terms of the experimental setup and is reflected in the mathematical framework of quantum mechanics. In this study, we shall treat not only the most important example of such incompatible properties --- the position and the momentum of a particle --- but also angle and angular momentum.

As has been argued by Bohr and Heisenberg \cite{heis1}, it is impossible to measure the position without loosing information about the momentum and vice versa. This property is embodied in the standard Heisenberg uncertainty relation and we will express it now in terms of Shannon measures of uncertainty.
\begin{align}\label{eur}
H^{(x)}+H^{(p)} > 1-\ln 2-\ln\left(\frac{\delta x\,\delta p}{h}\right).
\end{align}
To stress the physical meaning of this relation, we reinserted the Planck constant. It is clear that this inequality is not sharp since for large $\delta x\delta p$ the right hand side becomes negative. However, in the quantum regime, when the volume of phase space $\delta x\delta p$ does not exceed the Planck constant $h$, we obtain a meaningful limitation on the sum of the uncertainties in position and in momentum. When one uncertainty tends to zero, the other must stay above the limit. In the last example considered in the previous section the right hand side in (\ref{eur}) is equal to $1-\ln 2=0.307$, while the calculated value of the sum of two uncertainties was equal 1.223.

The proof of this inequality was given in \cite{ibbpl1} and it proceeds as follows. First, we use the integral form of the Jensen inequality for convex functions \cite{hlp,jen}. The function $\rho\ln\rho$ is a convex function. Therefore, the value of a function evaluated at the mean argument cannot exceed the mean value of the function so that the following inequality must hold:
\begin{align}\label{jen}
\langle\rho\ln\rho\rangle\ge \langle\rho\rangle\ln\langle\rho\rangle.
\end{align}
Upon substituting here the probability distribution, we obtain:
\begin{align}\label{jen1}
&\frac{1}{\delta x}\int_{(i-1/2)\delta x}^{(i+1/2)\delta x}\!\!\!dx\,\rho(x)\ln\rho(x)\nonumber\\
&\geq\left[\frac{1}{\delta x}\int_{(i-1/2)\delta x}^{(i+1/2)\delta x}\!\!\!dx\,\rho(x)\right]\,
\ln\left[\frac{1}{\delta x}\int_{(i-1/2)\delta x}^{(i+1/2)\delta x}\!\!\!dx\,\rho(x)\right],
\end{align}
or after some straightforward rearrangements and with the use of Eq.~(\ref{xbin})
\begin{align}\label{jen2}
-q_i\ln q_i\geq-\int_{(i-1/2)\delta x}^{(i+1/2)\delta x}\!\!\!dx\,\rho(x)\ln[\rho(x)\delta x].
\end{align}
This form of the inequality is more satisfactory from the physical point of view since $\rho(x)$ has the dimension of inverse length and the dimensional quantities should not appear under the logarithm. Adding the contributions from all the bins, we obtain:
\begin{align}\label{shx}
H^{(x)}\geq-\int_{-\infty}^{\infty}\!\!\!dx\,\rho(x)\ln[\rho(x)\delta x].
\end{align}
Applying the same reasoning to the momentum distribution, we arrive at:
\begin{align}\label{shp}
H^{(k)}\geq-\int_{-\infty}^{\infty}\!\!\!dk\,{\tilde\rho}(k)\ln[{\tilde\rho}(k)\delta k].
\end{align}
Adding these two inequalities, we get on the left hand side the sum of the uncertainties in position and in momentum as in (\ref{eur}). What remains is to establish a bound on the sum of the integrals appearing on the right hand side. This problem has a long history. More than half a century ago a bound has been conjectured by Hirschman \cite{hirsch} and Everett \cite{ever0,ever}. Actually, Hirschman proved only a weaker form of the inequality and Everett showed that the left hand side in (\ref{eur}) is stationary for Gaussian wave functions. The inequality was later proved by Bialynicki-Birula and Mycielski \cite{bbm} and independently by Beckner \cite{beck}. In this way we arrive at the entropic uncertainty relation (\ref{eur}). We shall give a detailed derivation of this inequality and its extensions in the next section.

The measure of uncertainties with the use of standard deviations requires a certain measure of the distance between events. Sometimes there is no sensible definition of a distance. The simplest example is the original application of the Shannon information entropy to text messages. The value of $H$ can be calculated for each text but there is no sensible measure of a distance between the letters in the alphabet. Important examples are also found in physics. The simplest case is a quantum particle moving on a circle. The configuration space is labeled by the angle $\varphi$ and the canonically conjugate variable is the angular momentum represented in quantum mechanics by the operator ${\hat M}=-i\hbar\partial/\partial\varphi$. Wave functions describing this system are integrable {\em periodic} functions $\psi(\varphi)$ or alternatively the coefficients $c_m$ in the Fourier expansion of $\psi(\varphi)$,
\begin{align}\label{fe}
\psi(\varphi)=\frac{1}{\sqrt{2\pi}}\sum_{m=-\infty}^\infty c_m\,e^{im\varphi}.
\end{align}
Even though formally $\varphi$ and ${\hat M}$ obey canonical commutation relations, these relations are not mathematically consistent because $\varphi$ cannot be treated as an operator --- multiplication by $\varphi$ produces a function that is not periodic. Therefore, the notion of a standard deviation of $\varphi$ can only be used approximately for states for which $\Delta\varphi$ is very small. Several, more or less natural methods, were introduced to deal with this problem. One of them is to use periodic functions of $\varphi$ instead of the angle itself (cf. \cite{reh} for a list of references). This leads to some uncertainty relations which are mathematically correct but physically less transparent. The use of entropic measures of uncertainties solves this problem. The uncertainty in position on the circle is defined as for position on an infinite line, except that now there is finite number of bins $N=2\pi/\delta\varphi$. Thus, the Shannon entropy of the angle is:
\begin{align}\label{csang}
H^{(\varphi)} = -\sum_{n=0}^{N-1} q_n\ln q_n,
\end{align}
where
\begin{align}\label{angbin}
q_n=\int_{n\delta\varphi}^{\left(n+1\right)\delta\varphi}\!\!d\varphi\,|\psi(\varphi)|^2.
\end{align}
According to the rules of quantum mechanics, the probability to find the value $\hbar m$ of angular momentum is $p_m=|c_m|^2$. Therefore, the Shannon entropy for the angular momentum is:
\begin{align}\label{csm}
H^{(M)} = -\sum_{m=-\infty}^{\infty}p_m\ln p_m.
\end{align}
The proof of the uncertainty relation for the angle and angular momentum entropies is much simpler than for the position and momentum. It will be given later in the more general case of the R\'enyi entropies. The uncertainty relation has the following form:
\begin{align}\label{euram}
H^{(\varphi)}+H^{(M)}\ge -\ln\frac{\delta\varphi}{2\pi},
\end{align}
or
\begin{align}\label{euram1}
H^{(\varphi)}+H^{(M)}\ge \ln N.
\end{align}
This inequality is saturated for every eigenstate of angular momentum. Then, the uncertainty in angular momentum vanishes and the uncertainty in position is exactly $\ln N$ because the probability to find the particle in a given bin is $1/N$.

\section{R\'enyi entropy}

The Shannon information entropy has been generalized by R\'enyi \cite{renyi}. The R\'enyi entropy is a one-parameter family of entropic measures that share with the Shannon entropy several important properties. Even though the R\'enyi entropy was introduced in 1960, its substantial use in physics is more recent --- it took place during the last decade (cf. references listed in the Introduction).

R\'enyi entropy $H_\alpha$ is defined by the following formula:
\begin{align}\label{renyi}
H_\alpha=\frac{1}{1-\alpha}\ln\left[\sum_k p_{k}^{\alpha}\right],
\end{align}
where $p_k$ is a set of probabilities and $\alpha$ is a positive number. R\'enyi in \cite{renyi} restricted $\alpha$ to positive values but we may consider, in principle, all values. Sometimes it is useful to consider only the values $\alpha>1$ since we may then introduce a conjugate positive parameter $\beta$ satisfying the relation
\begin{align}\label{mink}
\frac{1}{\alpha}+\frac{1}{\beta}=2.
\end{align}
In the limit, when $\alpha\to 1$, the R\'enyi entropy becomes the Shannon entropy. To see this, we have to apply the L'H\^opital rule to the definition (\ref{renyi}) at the singular point $\alpha=1$. Since the Shannon entropy is a special case of the R\'enyi entropy, we shall proceed, whenever possible, with proving various properties of the R\'enyi entropy, taking at the end the limit $\alpha\to 1$. This approach is particularly fruitful in the case of entropic uncertainty relations because the proofs are in a way more natural for the R\'enyi entropies than for the Shannon entropies.

Shannon entropy can be defined as a function $H(p_k)$ of the set of probabilities, obeying the following axioms:
\begin{enumerate}
\item{$H(p_1, p_2,\dots,p_n)\;{\rm is\;a\;symmetric\;function\;of\;its\;variables\;for}\; n = 2, 3,\dots$}
\item{$H(p, 1 - p)\;{\rm is\;a\;continuous\;function\;of}\; p\;{\rm for}\;\;0\leq p\leq 1$}
\item{$H(tp_1, (1 - t)p_1,p_2,\dots,p_n) = H(p_1, p_2,\dots,p_n) + p_1H(t, 1 -t)\;{\rm for}\;\;0\leq t\leq 1$}
\end{enumerate}
We presented the axioms in the form taken from R\'enyi \cite{renyi} which we liked more than the original axioms given by Shannon \cite{shan}. Sometimes an additional axiom is added to fix the scale (the base of the logarithm) of the function $H$. For example by requiring that $H(1/2,1/2)=1$ we obtain the entropy measured in bits (logarithm to the base 2).

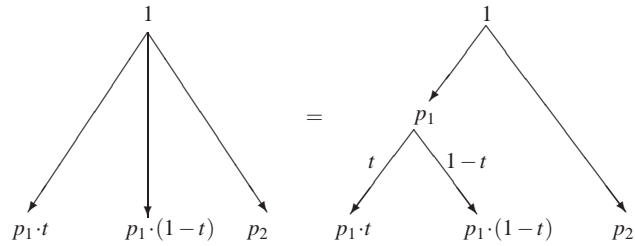
\begin{figure}
\hspace{1.0cm}
\unitlength=0.3mm
\begin{picture}(400, 140)(-50,0)
  \put(28,130){$1$}
  \put(30, 125){\vector(2, -3){53}}
  \put(30, 125){\vector(-2, -3){53}}
  \put(30, 125){\vector(0, -3){82}}
  \put(100,85){$=$}

  \put(-30,35){$p_1\!\cdot\!t$}
  \put(20,35){$p_1\!\cdot\!(1-t)$ }
  \put(74,35){$p_2$}
  \put(113,35){$p_1\!\cdot\!t$}
  \put(170,35){$p_1\!\cdot\!(1-t)$}
  \put(236,35){$p_2$}

  \put(128,65){$t$}
  \put(162,65){$1-t$}

  \put(178,130){$1$}
  \put(148, 82){\vector(3, -4){28}}
  \put(148, 86){$p_1$}
  \put(180, 128){\vector(-3, -4){25}}
  \put(148, 82){\vector(-3,-4){28}}
  \put(180, 128){\vector(3,-4){62}}
\end{picture}
\caption[Diagram]{Probability trees representing the main idea of the third axiom}\label{tree}
\end{figure}

The first two axioms are of a simple mathematical nature but the third axiom is of crucial importance. This axiom may be figuratively expressed as the invariance of the entropy with respect to breaking down the probability tree. This invariance rule is illustrated in Fig.~\ref{tree}. It leads, in particular, to the very important property that characterizes the physical entropy. The entropy is an extensive quantity --- for a composed system it must satisfy the additivity law. This law in terms of probabilities can be expressed as follows. Given two independent probability distributions $(p_1,p_2,\dots,p_n)$ and $(q_1,q_2,\dots,q_m)$, the entropy of the composed system must be equal to the sum of the entropies for the subsystems,
\begin{align}\label{add}
&H(p_1q_1,p_1q_2,\dots,p_1q_m,p_2q_1,\dots,p_2q_m,\dots,p_nq_1,\dots,p_nq_m)\nonumber\\
&=H(p_1,p_2,\dots,p_n)+H(q_1,q_2,\dots,q_m).
\end{align}
It turns out that by replacing the third axiom by the additivity law we substantially increase the set of allowed functions. The most prominent of these generalized entropies is the R\'enyi entropy. It is simple task to check that the R\'enyi entropy obeys the additivity law. Indeed, from the distributive property of the logarithm we obtain
\begin{align}\label{renyi1}
&\frac{1}{1-\alpha}\ln\left[\sum_{kl} (p_kq_l)^\alpha\right]=\frac{1}{1-\alpha}\ln\left[\sum_k p_k^\alpha \sum_lq_l^\alpha\right]\nonumber\\
&=\frac{1}{1-\alpha}\ln\left[\sum_k p_k^\alpha\right]+\frac{1}{1-\alpha}\ln\left[\sum_l q_l^\alpha\right].
\end{align}
Note that any linear combination (discrete or continuous) of  R\'enyi entropies with different values of $\alpha$ would obey the same additivity law. For example, we may take a sum of two terms
\begin{align}\label{renyi2}
\frac{1}{1-\alpha}\ln\left[\sum_k p_k^\alpha\right]+\frac{1}{1-\beta}\ln\left[\sum_k p_k^\beta\right].
\end{align}
A particular expression of this type was introduced in \cite{ibbur} and is very useful in the context of entropic uncertainty relation. It is obtained by requiring that the two parameters $\alpha$ and $\beta$ are constrained by the relation (\ref{mink}). In this case, they can be expressed in terms of a single variable: $\alpha=1/(1-s),\;\beta=1/(1+s)$. The new symmetrized entropy ${\cal H}_s$ is a one parameter average of two R\'enyi entropies,
\begin{align}\label{brack}
{\cal H}_s=\frac{1}{2}(1-\frac{1}{s})\ln\left[\sum_k p_k^{1/(1-s)}\right]
+\frac{1}{2}(1+\frac{1}{s})\ln\left[\sum_k p_k^{1/(1+s)}\right].
\end{align}
Since ${\cal H}_s$ is a symmetric function of $s$, it is sufficient to consider only the positive values $0\leq s\leq 1$. The Shannon entropy is recovered in the limit when $s\to 0$.

The R\'enyi entropy is a decreasing function of $\alpha$. For a given set of probabilities it attains its maximal value at $\alpha=0$. The symmetrized R\'enyi entropy attains its maximal value at $s=0$. Starting from this value, as a symmetric function of $s$, it drops down in both directions.

\subsection{R\'enyi entropy and the uncertainty relations}

Our goal is now to generalize the uncertainty relations to the R\'enyi entropies. In this section we shall consider only the measurements of position and momentum. We have already defined the Shannon information entropies that contain information about the precision of measurements. The definitions of the uncertainty in position and momentum, as measured by the R\'enyi entropy, are obvious:
\begin{align}\label{renyixk}
H_\alpha^{(x)}&=\frac{1}{1-\alpha}\ln\left[\sum_{i=-\infty}^\infty q_i^\alpha\right],\\
H_\beta^{(k)}&=\frac{1}{1-\beta}\ln\left[\sum_{j=-\infty}^\infty p_j^\beta\right],
\end{align}
where the probabilities $q_i$ and $p_j$ are given by the formulas (\ref{xbin}) and (\ref{kbin}). The reason, why we have chosen the R\'enyi entropies with different parameters will become clear later. For definiteness, we shall assume at this point that $\alpha\ge\beta$ (later this restriction will be lifted). This means that according to (\ref{mink}) $1\le\alpha<\infty\,$ and $1/2<\beta\le 1$. Therefore, $z^\alpha$ is a convex function, while $z^\beta$ is concave. These properties enable us to use the Jensen inequality in the same way as we did before, and arrive at:
\begin{subequations}
\begin{align}
\left(\frac{1}{\delta x}
\int_{\left(i-1/2\right)\delta x}^{\left(i+1/2\right)\delta x}dx\,\rho(x)\right)^{\alpha}\leq\frac{1}{\delta x}\int_{\left(i-1/2\right)\delta x}^{\left(i+1/2\right)\delta x}\!dx
\left[\rho(x)\right]^{\alpha},\label{Q95}\\
\frac{1}{\delta k}\int_{\left(j-1/2\right)\delta k}^{\left(j+1/2\right)\delta k}\!dk
\left[\tilde{\rho}(k)\right]^\beta\leq\left(\frac{1}{\delta k}\int_{\left(j-1/2\right)\delta k}^{\left(j+1/2\right)\delta k}\!dk\tilde{\rho}
(k)\right)^{\beta}.\label{Q96}
\end{align}
\end{subequations}
Summing (\ref{Q95}) and (\ref{Q96}) over the indices $i$ and $j$ we obtain:
\begin{subequations}
\begin{align}
\left(\delta x\right)^{1-\alpha}\sum_{i=-\infty}^{\infty}q_i^{\alpha}
\leq\int_{\mathbb{R}}\!dx\left[\rho(x)\right]^{\alpha},\label{Q97}\\
\int_{\mathbb{R}}\!dk\left[\tilde{\rho}(k)\right]^{\beta}\leq\left(\delta k\right)^{1-\beta}\sum_{j=-\infty}^{\infty}p_j^{\beta}.\label{Q98}
\end{align}
\end{subequations}
At this point we invoke a powerful Babenko-Beckner inequality for $(p,q)$ norms of a function and its Fourier transform. This inequality was proved for restricted values of $\alpha$ and $\beta$ by Babenko \cite{bab} and for all values by Beckner \cite{beck}. In our notation, this inequality reads:
\begin{align}
\left(\int_{\mathbb{R}}dx\,\left[\rho(x)\right]^{\alpha}\!\right)^{1/\alpha}\leq n\left(\alpha,\beta\right)\left(\int_{\mathbb{R}}\!dk
\left[\tilde{\rho}(k)\right]^{\beta}\right)^{1/\beta},\label{Q86}
\end{align}
where
\begin{align} n\left(\alpha,\beta\right)=\left(\frac{\alpha}{\pi}\right)^{-1/2\alpha}
\left(\frac{\beta}{\pi}\right)^{1/2\beta}.\label{Q87}
\end{align}
This inequality enables us to put together the formulas (\ref{Q97}) and (\ref{Q98}). To this end, we raise to the power $1/\alpha$ both sides of (\ref{Q97}). Next, we raise to the power of $1/\beta$ both sides of (\ref{Q98}) and multiply the resulting inequality by $n\left(\alpha,\beta\right)$. In this way we can obtain the following chain of inequalities:
\begin{align}\label{chain}
&\left[\left(\delta x\right)^{1-\alpha}\sum_{i=-\infty}^{\infty}q_i^{\alpha}\right]^{1/\alpha}
\leq\left(\int_{\mathbb{R}}dx\,\left[\rho(x)\right]^{\alpha}\!\right)^{1/\alpha}\nonumber\\
&\leq n\left(\alpha,\beta\right)\left(\int_{\mathbb{R}}\!dk
\left[\tilde{\rho}(k)\right]^{\beta}\right)^{1/\beta}
\leq n\left(\alpha,\beta\right)\left[\left(\delta k\right)^{1-\beta}\sum_{j=-\infty}^{\infty}p_j^{\beta}\right]^{1/\beta}.
\end{align}
Now, we take only the first and the last term of this chain and multiply both sides of the resulting inequality by $(\alpha/\pi)^{1/2\alpha}$ to obtain:
\begin{align}\label{chain1}
&\left[\left(\delta x\right)^{1-\alpha}\sqrt{\frac{\alpha}{\pi}}\sum_{i=-\infty}^{\infty}q_i^{\alpha}\right]^{1/\alpha}
\leq\left[\left(\delta k\right)^{1-\beta}\sqrt{\frac{\beta}{\pi}}\sum_{j=-\infty}^{\infty}p_j^{\beta}\right]^{1/\beta}.
\end{align}
Next, we evaluate the logarithms of both sides and multiply the result by a positive number $\alpha/(\alpha-1)$. After a rearrangement of terms, with the use of the relationship (\ref{mink}), we arrive at the final form of the uncertainty relation (with Planck's constant reinserted) between position and momentum in terms of the R\'enyi entropies \cite{ibbur}:
\begin{align}\label{Q100}
H_\alpha^{(x)}+H_\beta^{(p)}\geq
-\frac{1}{2}\left(\frac{\ln\alpha}{1-\alpha}+\frac{\ln\beta}{1-\beta}\right)-\ln 2
-\ln\frac{\delta x\delta p}{h}.
\end{align}
We can lift now the restriction that $\alpha\geq\beta$. Owing to a symmetry between a function and its Fourier transform we can write the same inequality (\ref{Q100}) but with $x$ and $p$ interchanged. After all, it does not make any difference from the mathematical point of view, what is called $x$ and what is called $p$, as long as the corresponding wave functions are connected by the Fourier transformation.  Therefore, we can take the average of (\ref{Q100}) and its counterpart with $x$ and $p$ interchanged and obtain the following uncertainty relation for symmetrized entropies:
\begin{align}\label{Q100s}
{\cal H}_s^{(x)}+{\cal H}_s^{(p)}\geq
\frac{1}{2}\ln(1-s^2)+\frac{1}{2s}\ln\frac{1+s}{1-s}-\ln 2
-\ln\frac{\delta x\delta p}{h}.
\end{align}
This relation is more satisfactory than (\ref{Q100}) because the uncertainties of both quantities $x$ and $p$ are measured in the same way. In the limit, when $s\to 0$ we obtain the uncertainty relation (\ref{eur}) for the Shannon entropies. The inequality (\ref{Q100s}) (or its asymmetric counterpart (\ref{Q100})) is the best known uncertainty relation for the entropies that takes into account finite resolution of measurements. However, this inequality is obviously not a sharp one. For $\delta x\delta p/h$ only slightly larger than 1 the right hand side of (\ref{Q100}) becomes negative while the left hand side is by definition positive. For finite values of $\delta x\delta p/h$ there must exist a better lower bound. Since the Babenko-Beckner inequality (\ref{Q86}) cannot be improved (it is saturated by Gaussian functions), we conclude that Jensen inequalities (\ref{Q95}, \ref{Q96}) used together are not optimal. Of course, each of them can be separately saturated but when they are used jointly one should somewhere include the information about the Fourier relationship between the functions. An improvement of the uncertainty relations (\ref{Q100}) or (\ref{Q100s}) is highly desired but it does not seem to be easy. A recent attempt \cite{wilk} failed completely (see \cite{comment}).

\section{Uncertainty relations: from discrete to continuous}

In this section we describe various types of uncertainty relations that depart from the original Heisenberg relations for position and momentum much further than those described so far. We shall also comment on different mathematical tools that are used in proving these relations.

\subsection{The uncertainty relations for $N$-level systems}

Pure states of an $N$-level system are described in quantum mechanics by normalized vectors in the $N$-dimensional Hilbert space $\mathcal{H}_{N}$. We shall use the Dirac notation denoting vectors by  $|\psi\rangle$ and their scalar products by  $\langle\psi_1|\psi_2\rangle$. In this notation, the normalization condition reads $\langle\psi|\psi\rangle=1$. Let us choose two orthonormal bases $\left\{ |a_i\rangle\right\} $ and $\left\{ |b_j\rangle\right\} $, where $i,j=1,\ldots,N$. We can then represent each vector $|\psi\rangle$ in two ways:
\begin{align}
|\psi\rangle=\sum_{i=1}^{N}\langle a_i|\psi\rangle|a_i\rangle
=\sum_{j=1}^{N}\langle b_j|\psi\rangle|b_j\rangle.\label{Q46}
\end{align}
The squares of the absolute values of the complex expansion coefficients are interpreted in quantum mechanics as the probabilities, say $q_i$ and $p_j$, to find the states represented by the vectors $|a_i\rangle$ and $|b_j\rangle$, when the system is in the state $|\psi\rangle$,
\begin{align}
q_i=\left|\langle a_i|\psi\rangle\right|^{2},\qquad p_j=\left|\langle b_j|\psi\rangle\right|^{2}.\label{Q47}
\end{align}
The normalization of vectors and the orthonormality of the bases guarantees that the probabilities  $q_i$ and $p_j$ sum up to 1.

\subsubsection{Deutsch inequality}

Having at our disposal two sets of probabilities $\{q_i\}$ and $\{p_j\}$ we can construct two Shannon entropies:
\begin{align}\label{Q48}
H^{(a)}=-\sum_{i=1}^Nq_i\ln q_i,\quad H^{(b)}=-\sum_{j=1}^Np_j\ln p_j.
\end{align}
Since we aim at finding some uncertainty relation we consider the sum of these entropies:
\begin{align}
H^{(a)}+H^{(b)}=\sum_{i=1}^{N}\sum_{j=1}^{N}q_ip_jQ_{ij},\label{Q50}
\end{align}
where
\begin{align} Q_{ij}=-\ln\left(\langle a_i|\psi\rangle\langle\psi|a_i\rangle\right)-\ln\left(\langle b_j|\psi\rangle\langle \psi|b_j\rangle\right).\label{Q50a}
\end{align}
In 1983 Deutsch \cite{d} found a lower bound on the sum (\ref{Q50}). He observed that for a normalized vector $|\psi\rangle$ each $Q_{ij}$ is nonnegative. Therefore, by finding the minimal value of all $Q_{ij}$'s  we determine a lower bound on the sum (\ref{Q50}). We shall present here an expanded version of the Deutsch analysis. For each pair $(i,j)$, the minimal value of $Q_{ij}$ is attained for some function $\psi_{ij}$. This function is found by varying $Q_{ij}$ with respect to the wave function $\psi$ with the normalization constraint $\langle\psi|\psi\rangle=1$.
\begin{align}
\frac{\delta}{\delta|\psi_{ij}\rangle}\left[Q_{ij}+\kappa\langle \psi_{ij}|\psi_{ij}\rangle\right]=0,\label{Q51}
\end{align}
where $\kappa$ is a Lagrange multiplier. In this way we arrive at the equation for the vector $|\psi_{ij}\rangle$:
\begin{align}
|\psi_{ij}\rangle=\frac{1}{\kappa}\left(\frac{|a_i\rangle}{\langle \psi_{ij}|a_i\rangle}+\frac{|b_j\rangle}{\langle\psi_{ij}|b_j\rangle}\right).\label{Q52}
\end{align}
Taking the scalar product of this equation with $\langle\psi_{ij}|$, we obtain from the normalization condition that $\kappa=2$. To find the solution of (\ref{Q52}) we multiply this equation by the adjoined vectors $\langle a_i|$ and $\langle b_j|$. In this way we obtain the following algebraic equations for two complex variables $a=\langle\psi_{ij}|a_i\rangle$ and $b=\langle\psi_{ij}|b_j\rangle$ with one fixed complex parameter $\langle a_i|b_j\rangle$.
\begin{align}\label{ab}
|a|^2=\frac{1}{2}+\frac{a\langle a_i|b_j\rangle}{2b},\quad|b|^2=\frac{1}{2}+\frac{b\langle b_j|a_i\rangle}{2a}.
\end{align}
The last term in each equation must be real since the rest is real. Next, we divide both sides of the first equation by those of the second. After some simple rearrangements, we arrive at:
\begin{align}\label{ab1}
\frac{|a|^2}{|b|^2}-1=\frac{a}{b}\langle a_i|b_j\rangle-\frac{a^*}{b^*}\langle a_i|b_j\rangle^*.
\end{align}
Since both terms on the right hand side are real, they cancel out and we obtain $|a|=|b|$. Inserting this into any of the two equations (\ref{ab}), we obtain
\begin{align}\label{ab2}
|a|^2=\frac{1}{2}\left[1\pm|\langle a_i|b_j\rangle|\right].
\end{align}
The choice of the minus sign gives the maximal value of $Q_{ij}$ but only in the subspace spanned by the vectors $|a_i\rangle$ and $|b_j\rangle$. Choosing the plus sign, we obtain
\begin{align}
|\psi_{ij}\rangle=\frac{\exp(i\phi)}{\sqrt{2\left(1+|\langle a_i|b_j\rangle|\right)}}\left[|a_i\rangle+\exp\left(-i\arg\langle a_i|b_j\rangle\right)|b_j\rangle\right],\label{Q53}
\end{align}
where $\exp(i\phi)$ is an arbitrary phase factor.
We can insert this solution into (\ref{Q50}), and using the fact that this solution is minimizer, we obtain the inequality:
\begin{align}
H^{(a)}+H^{(b)}\ge-2\sum_{i=1}^N\sum_{j=1}^Nq_ip_j
\ln\left[\frac{1}{2}\left(1+|\langle a_i|b_j\rangle|\right)\right].\label{Q54}
\end{align}
This inequality will still hold if we replace each modulus $|\langle a_i|b_j\rangle|$ by the maximal value $C_B$,
\begin{align}
C_{B}=\sup_{(i,j)}|\langle a_i|b_j\rangle|.\label{Q56}
\end{align}
After this replacement, we can do summations over $i$ and $j$ and arrive at the Deutsch result \cite{d}:
\begin{align}
H^{(a)}+H^{(b)}\ge-2\ln\left[\frac{1}{2}\left(1+C_{B}\right)\right].\label{Q55}
\end{align}

\subsubsection{Maassen-Uffink inequalities}\label{E2}

Even though the bound in the Deutsch uncertainty relation is saturated when the two bases have common vectors, there is room for improvement. Of course, the improved inequality must coincide with  (\ref{Q55}) when it is saturated. An improved relation was conjectured by Kraus \cite{kraus} and it has the form:
\begin{align}
H^{(a)}+H^{(b)}\ge-2\ln C_{B},\label{Q57}
\end{align}
This inequality was later proved by Maassen and Uffink \cite{mu}. We shall present a different version of the proof based on the R\'enyi entropy. The mathematical tools needed to prove the inequality (\ref{Q57}) are the same as those used for the entropies with continuous variables. At the bottom of every uncertainty relation there is some mathematical theorem. In the case of the Maassen-Uffink relation this role is played by the Riesz theorem \cite{rs,riesz} which says that for every $N$-dimensional complex vector $X$ and a unitary transformation matrix $\hat{T}$ with coefficients $t_{ji}$, the following inequality between the norms is valid:
\begin{align} c^{1/\mu}||X||^\mu\leq c^{1/\nu}||{\hat{T}}X||^\nu,\label{Q58}
\end{align}
where the constant $c=\sup_{(i,j)}\left|t_{ji}\right|$ and the coefficients $\mu$ and $\nu$ obey the relation \begin{align}
\frac{1}{\mu}+\frac{1}{\nu}=1,\qquad1\leq\nu\leq2.\label{Q59}
\end{align}
The norms are defined as usual,
\begin{align}
||X||^\mu=\left[\sum_k|x_k|^\mu\right]^{1/\mu},
\end{align}
but we do not use the traditional symbols $(p,q)$ for the norms since this would interfere with our usage of $p$ and $q$ to denote the probabilities.

To prove the inequality (\ref{Q57}) we shall take $x_i=\langle a_i|\psi\rangle$ and
\begin{align}\label{Q59a}
t_{ji}=\langle b_j|a_i\rangle.
\end{align}
>From the resolution of the identity ($\sum_i|a_i\rangle\langle a_i|=1$), we obtain:
\begin{align}
\sum_{i=1}^{N}t_{ji}x_i=\sum_{i=1}^{N}\langle b_j|a_i\rangle\langle a_i|\psi\rangle
=\langle b_j|\psi\rangle.\label{Q60}
\end{align}
Using the definitions (\ref{Q47}), we can rewrite the inequality (\ref{Q58}) in the form:
\begin{align}
c^{1/\mu}\left[\sum_{j=1}^{N}q_j^{\mu/2}\right]^{1/\mu}\leq c^{1/\nu}\left[\sum_{i=1}^{N}p_i^{\nu/2}\right]^{1/\nu},\label{Q61}
\end{align}
where
\begin{align}\label{c}
c=\sup_{(i,j)}|t_{ij}|=\sup_{(i,j)}|\langle a_i|b_j\rangle|=C_{B}.
\end{align}
The parameters $\mu$ and $\nu$ differ by a factor of 2 from the parameters $\alpha$ and $\beta$ appearing in (\ref{mink}}),
\begin{align}
\mu=2\alpha,\;\nu=2\beta.\label{Q62}
\end{align}
Next, we take the logarithm of both sides of the inequality (\ref{Q61}) and with the use of Eqs.~(\ref{Q62}), we arrive at the uncertainty relation for the R\'enyi entropies:
\begin{align}
H_\alpha^{(a)}+H_\beta^{(b)}\geq-2\ln C_{B}.\label{Q63}
\end{align}
When two observables associated with the two bases have a common vector then $C_B=1$ and the right hand side in the last inequality vanishes. Thus, this uncertainty relation becomes empty. In the limit $\alpha\to 1$, $\beta\to 1$ this inequality reduces to the Maassen-Uffink result (\ref{Q57}).

\subsubsection{Discrete Fourier transform}\label{mut}

In previous subsections we dealt with two orthonormal vector bases $\left\{ |a_{k}\rangle\right\} $ and $\left\{ |b_{m}\rangle\right\} $. We can expand every vector of the first basis in the second basis,
\begin{align}
|a_i\rangle=\sum_{j=1}^Nt_{ji}|b_j\rangle,\label{Q64}
\end{align}
where $t_{ji}$ are the coefficients of the unitary transformation (\ref{Q59a}),
and for each $|a_i\rangle$
\begin{align}
1=\langle a_i|a_i\rangle=\sum_{j=1}^N|t_{ji}|^{2}\leq N\sup_{(j)}|t_{ji}|^{2}\leq NC_{B}^{2}.\label{Q67}
\end{align}
It follows from this inequality that $C_{B}\geq 1/\sqrt{N}$. There are bases, called mutually unbiased bases, for which the sum (\ref{Q63}) not only has the lowest possible value but the moduli of the matrix elements are {\em all} the same,
\begin{align}\label{mub}
|\langle a_i|b_j\rangle|=\frac{1}{\sqrt{N}}.
\end{align}
One can check by a direct calculation that for the observables associated with two mutually unbiased bases, for every state represented by a basis vector, either $|a_i\rangle$ or $|b_j\rangle$, the sum of the R\'enyi entropies saturates the lower bound,
\begin{align}
H_\alpha^{(a)}+H_\beta^{(b)}=\ln N.\label{Q69}
\end{align}

Mutually unbiased bases were introduced by Schwinger \cite{sch}. More recently they were studied in more detail \cite{beng} and were shown to be useful in the description of entanglement and other fundamental properties of quantum systems. In the fourth chapter we shall say more about the mutually unbiased bases. The most important example of mutually unbiased bases is present in the discrete Fourier transformation. In this case, the two bases are related by the unitary transformation with the coefficients:
\begin{align}
f_{kl}=\frac{1}{\sqrt{N}}\exp\left[\frac{2\pi i\,kl}{N}\right],\label{Q68}
\end{align}
which satisfy the conditions (\ref{mub}).

The physical interpretation of the inequality (\ref{Q69}) is similar to the Heisenberg uncertainty relation. Two observables $A$ and $B$, characterized by the bases ${|a_i\rangle}$ and ${|b_j\rangle}$ cannot have simultaneously sharp values since they do not have common basis vectors. Moreover, every state described by a basis vector has the sum of the uncertainties equal to the same value $1/\sqrt{N}$. The observables $A$ and $B$ are finite-dimensional analogs of canonically conjugate physical quantities.

\subsection{Shannon and R\'enyi entropies for continuous distributions}\label{E4}

In a rather crude manner we can say that the entropies for continuous distributions of probability are obtained by replacing the sums in the expressions (\ref{cs}) and (\ref{renyi}) by the integrals. We have already encountered such expressions, (Eqs.~(\ref{shx}) and (\ref{shp})), in our discussion of the Shannon entropies for position and momentum. Naively, we would like to define the Shannon entropy for a continuous probability distribution $\rho(X)$ in the form:
\begin{align}
H^{(X)}\,{\overset{?}=}\,-\int dX\rho(X)\ln\rho(X),\label{Q20}
\end{align}
where $X$ is some random variable. However, in most cases we encounter a problem since $X$ is a dimensional quantity and the result will depend on the unit that is used to measure $X$. In order to obtain a correct formula, let us start with the original Shannon definition for a discrete set of probabilities (\ref{cs}). Next, we insert in this formula the discrete probabilities $p_k$ derived from some continuous distribution. For example, let us choose the distribution function for position $\rho(x)$. If $\rho(x)$ does not change appreciably over the distance $\delta x$, the Shannon entropy defined by Eqs.~(\ref{xbin}) and (\ref{csx}) can be approximated by,
\begin{align}\label{shxa}
H^{(x)}\approx-\sum_{k=-\infty}^\infty\delta x\,\rho(x_k)\ln[\rho(x_k)\delta x],
\end{align}
where $x_k=k\delta x$.

We can write the right hand side as a sum of two terms
\begin{align}\label{shx1}
-\sum_{k=-\infty}^\infty\delta x\,\rho(x_k)\ln[\rho(x_k)L]-\ln\frac{\delta x}{L},
\end{align}
where $L$ is some fixed unit of length. The first term is a Riemann sum and in the limit, when $\delta x\to 0$, it tends to the following integral:
\begin{align}\label{shx2}
S^{(x)}=-\int_{\mathbb{R}}\!dx\,\rho(x)\ln[\rho(x)L].
\end{align}
This integral may be called the entropy of the continuous distribution $\rho(x)$ or the continuous entropy.

The second term $-\ln(\delta x/L)$ measures the difference between $H^{(x)}$ and $S^{(x)}$. This difference must tend to $\infty$ because the information measured by $H^{(x)}$ grows indefinitely when $\delta x\to 0$, while $S^{(x)}$ remains finite.

The same reasoning leads to the following definition of the R\'enyi entropy for a continuous distribution
\begin{align}\label{rx}
S_\alpha^{(x)}=\frac{1}{1-\alpha}\ln\left(\int_{\mathbb{R}}\!dx\,[\rho(x)]^\alpha L^{\alpha-1}\right).
\end{align}
The difference between the R\'enyi entropies $H_\alpha^{(x)}$ and $S_\alpha^{(x)}$ does not depend on $\alpha$ and is, therefore, the same as for the Shannon entropies.

\subsection{Uncertainty relations for continuous entropies}

We can formulate now uncertainty relations in terms of the R\'enyi and Shannon entropies for continuous variables, although the operational meaning of these relations is not as clear as for discrete bins. In the proof we shall make use of the Babenko-Beckner inequality (\ref{Q86}). This inequality may, at first, seem to violate the requirement that it should be invariant under a change of the physical units used to measure the probability distributions $\rho(x)$ and ${\tilde\rho}(k)$. However, it turns out that this inequality is dimensionally correct. To see this, let us assume that we measure position using the unit $L$. It follows from the definition of the Fourier transform (\ref{ftk}) that to keep this relation invariant, we have to choose $1/L$ as the unit of $k$. Now, we multiply the left hand side of (\ref{Q86}) by 1 written in the form $L^{2-1/\alpha-1/\beta}$. After distributing the powers of $L$, we arrive at the inequality
\begin{align}\label{bbl}
\left(\int_{\mathbb{R}}dx\,\left[\rho(x)L\right]^{\alpha}\!/L\right)^{1/\alpha}\leq n\left(\alpha,\beta\right)\left(\int_{\mathbb{R}}\!dk\,
\left[\tilde{\rho}(k)/L\right]^{\beta}L\right)^{1/\beta}.
\end{align}
This inequality is not only explicitly dimensionally correct but it also shows its invariance under a change of the unit. Taking the logarithm of both sides of this inequality and using the relation (\ref{mink}), we arrive at the uncertainty relation for the continuous R\'enyi entropies,
\begin{align}\label{cer}
S_\alpha^{(x)}+S_\beta^{(k)}&=\frac{1}{1-\alpha}\ln\left(\int_{\mathbb{R}}\!\!dx\,[\rho(x)]^\alpha L^{\alpha-1}\right)+\frac{1}{1-\beta}\ln\left(\int_{\mathbb{R}}\!\!dk\,[{\tilde\rho}(k)]^\beta L^{1-\beta}\right)\nonumber\\
&\ge-\frac{1}{2}\left(\frac{1}{1-\alpha}\ln\frac{\alpha}{\pi}
+\frac{1}{1-\beta}\ln\frac{\beta}{\pi}\right).
\end{align}
Due to the relation
\begin{align}\label{mink1}
\frac{\alpha}{1-\alpha}+\frac{\beta}{1-\beta}=0,
\end{align}
the uncertainty relation is invariant under a rescaling of the wave function. In the limit, when both $\alpha$ and $\beta$ tend to 1, we obtain the uncertainty relation for the Shannon entropies proved in \cite{bbm} and in \cite{beck},
\begin{align}\label{shc1}
S^{(x)}+S^{(k)}=-\int_{\mathbb{R}}\!\!dx\,\rho(x)\ln\rho(x)
-\int_{\mathbb{R}}\!\!dk\,{\tilde\rho}(k)\ln{\tilde\rho}(k)\ge 1+\ln\pi.
\end{align}
In evaluating this limit by the L'H\^ospital rule we assumed that the wave function is normalized,
\begin{align}\label{norm1}
\int_{\mathbb{R}}\!\!dx\,|\psi(x)|^2 = 1.
\end{align}
After an arbitrary rescaling of the wave function, the uncertainty relation takes on the form:
\begin{align}\label{shc2}
S^{(x)}+S^{(k)}=-\int_{\mathbb{R}}\!\!dx\,\rho(x)\ln\rho(x)
-\int_{\mathbb{R}}\!\!dk\,{\tilde\rho}(k)\ln{\tilde\rho}(k)\ge N^2(1+\ln\pi-4\ln N),
\end{align}
where $N$ is the norm of the wave function. We dropped in (\ref{shc1}) the scale factor $L$ because the sum of the two integrals is invariant under a change of units.

\subsubsection{Gaussian wave functions}

Gaussian wave functions play a distinguished role in uncertainty relations. They saturate the standard uncertainty relations expressed in terms of standard deviations. They also saturate the uncertainty relations (\ref{cer}) expressed in terms of the R\'enyi entropies for continuous variables. Thus, unlike their discrete counterparts (\ref{eur}) or (\ref{Q100}), the continuous uncertainty relations are sharp. Gaussians provide also a direct link between the Shannon entropies for continuous variables and the uncertainties measured by the standard deviation. Namely, the Gaussian distribution function gives the maximal value of the Shannon entropy subjected to the conditions of normalization and a given value $\sigma_{x}$ of the standard deviation. To prove this property we shall search for the maximum of the functional:
\begin{align} \int_{\mathbb{R}}dx\,\left[-\rho(x)\ln\rho(x)+\lambda x^{2}\rho(x)+\mu\rho(x)\right],\label{Q25}
\end{align}
where $\lambda$ and $\mu$ are two Lagrange multipliers introduced to satisfy the constraints:
\begin{subequations}\label{Q26}
\begin{align}
\int_{\mathbb{R}}dx\,\rho(x)&=1,\\
\int_{\mathbb{R}}dx\,x^{2}\rho(x)&=\sigma_{x}^{2}.
\end{align}
\end{subequations}
Varying the functional (\ref{Q25}) with respect to $\rho(x)$ we obtain the expression:
\begin{align}
-1-\ln\rho(x)+\lambda x^{2}+\mu.\label{Q28}
\end{align}
>From the requirement that this expression vanishes, we obtain a Gaussian distribution. Finally, we determine the Lagrange multipliers $\lambda$ and $\mu$ by imposing the constraints (\ref{Q26}), and we obtain:
\begin{align} \rho(x)=\frac{1}{\sqrt{2\pi}\sigma_{x}}e^{-\left(x^{2}/2\sigma_{x}^{2}\right)}.\label{Q29}
\end{align}
This extremum is a true maximum since the entropy is a strictly concave functional of $\rho(x)$ \cite{bbm}.

\section{Generalizations and extensions}

In this chapter we present a generalization of the entropic uncertainty relations to cover mixed states and an extension of these relations to angular variables. We shall also introduce uncertainty relations in phase space. Next, we return to the problem of mutually unbiased bases, mentioned already in (\ref{mut}). Finally, we show a connection between the entropic uncertainty relations and the logarithmic Sobolev inequalities. Other extensions of mathematical nature of the entropic uncertainty relations were recently published by Zozor et. al. \cite{zozor}.

\subsection{Uncertainty relations for mixed states}

Let us take a set of normalized wave functions $\psi_i(x)$ that determine the probability distributions $\rho_i(x)$. Using these distributions we may define the probability distribution for a mixed state:
\begin{align}
\rho_{\rm mix}(x)=\sum_i\lambda_i\rho_i(x).\label{Ad1}
\end{align}
The sum may be finite or infinite. Alternatively, we may start from the density operator ${\hat\rho}$ and define the probability density as the diagonal matrix element $\rho_{\rm mix}(x)=\langle x|{\hat\rho}|x\rangle$. The positive coefficient $\lambda_i$ determines the probability with which the state described by $\psi_i(x)$ enters the mixture. The normalization condition ${\rm Tr}\{\hat\rho\}=1$, or the basic property of probabilities, require that
\begin{align}
\sum_i\lambda_i=1.\label{Ad2}
\end{align}
For each $\psi_i(x)$ we may introduce its Fourier transform $\tilde{\psi}_i(k)$ and then define the probability distribution $\tilde{\rho}_i(k)$ in momentum space.
\begin{align}
\tilde{\rho}_{\rm mix}(k)=\sum_i\lambda_i\tilde{\rho}_i(k),\label{Ad3}
\end{align}
This function can also be viewed as the momentum representation of the density operator, $\tilde{\rho}_{\rm mix}(k)=\langle k|{\hat\rho}|k\rangle$.

We begin with the following sum of Babenko-Beckner inequalities (\ref{Q86}) with wave functions $\psi_i(x)$ and $\tilde{\psi}_i(k)$ and with some weights $\lambda_i$:
\begin{align} \sum_i\lambda_i\left(\int_{\mathbb{R}}dx
\left|\rho_i(x)\right|^{\alpha}\right)^{1/\alpha}\leq n\left(\alpha,\beta\right)\sum_i\lambda_i
\left(\int_{\mathbb{R}}dk\left|\tilde{\rho}_i(k)\right|^{\beta}\right)^{1/\beta}.\label{Ad8}
\end{align}
To find the uncertainty relation for the mixed state we shall use now the Minkowski inequality \cite{hlp} as was done in \cite{ibbur}. For $\alpha\geq 1$ and for an arbitrary set of nonnegative functions $f_i(x)$ the Minkowski inequality reads:
\begin{align}
\left(\int_{\mathbb{R}}dx\left|\sum_i f_i(x)\right|^\alpha\right)^{1/\alpha}
\leq\sum_i\left(\int_{\mathbb{R}}dx\left|f_i(x)\right|^\alpha\right)^{1/\alpha}.\label{Ad4}
\end{align}
Since for $\alpha\geq 1$ we have $\beta\leq1$, the Minkowski inequality gets inverted. Choosing another set of nonnegative functions $g_i(k)$ we obtain:
\begin{align}\label{Ad5}
\sum_i\left(\int_{\mathbb{R}}dk\left|g_i(k)\right|^\beta\right)^{1/\beta}
\leq\left(\int_{\mathbb{R}}dk\left|\sum_i g_i(k)\right|^\beta\right)^{1/\beta}.
\end{align}
Substituting now $f_i(x)=\lambda_i\rho_i(x)$ in (\ref{Ad4}) and $g_i(k)=\lambda_i\tilde{\rho}_i(k)$ in (\ref{Ad5}) we obtain two inequalities involving the densities for mixed states:
\begin{align}\label{Ad6} \left(\int_{\mathbb{R}}dx\left|\rho_{\rm mix}(x)\right|^\alpha\right)^{1/\alpha}
\leq\sum_i\lambda_i\left(\int_{\mathbb{R}}dx\left|\rho_i(x)\right|^{\alpha}\right)^{1/\alpha}.
\end{align}
and
\begin{align}\label{Ad7}
\sum_i\lambda_i\left(\int_{\mathbb{R}}dk\left|\tilde{\rho}_i(k)\right|^\beta\right)^{1/\beta}
\leq\left(\int_{\mathbb{R}}dk\left|\tilde{\rho}_{\rm mix}(k)\right|^\beta\right)^{1/\beta}.
\end{align}
Putting together the inequalities (\ref{Ad8}), (\ref{Ad6}), and (\ref{Ad7}) we obtain the generalization of the Babenko-Beckner inequality for mixed states:
\begin{align} \left(\int_{\mathbb{R}}dx\left[\rho_{\rm mix}(x)\right]^{\alpha}\right)^{1/\alpha}\leq n\left(\alpha,\beta\right)\left(\int_{\mathbb{R}}dk\left[\tilde{\rho}_{\rm mix}(k)\right]^{\beta}\right)^{1/\beta}.\label{Ad9}
\end{align}
>From this inequality we can prove the uncertainty relations (\ref{Q100}) and (\ref{Q100s}) for mixed states in the same manner as we have done it for pure states.

\subsection{Uncertainty relations for angles and angular momenta}

In (\ref{mom}) we gave the uncertainty relation for angle $\varphi$ and angular momentum $M$ in terms of the Shannon entropy. Now we are going to prove this relation in a more general case of the R\'enyi entropy. To this end we shall repeat and extend the previous definitions.

\subsubsection{The uncertainty relation for a particle on a circle}

We have discussed this problem already in Sec. \ref{mom} and now we shall provide the proofs of the uncertainty relations. The proof of the uncertainty relation (\ref{euram}) in the general case of the R\'enyi entropy will be given in two steps.

First, we quote the Young-Hausdorff inequality for the Fourier series \cite{ibbur}:
\begin{align} \left(\int_{0}^{2\pi}d\varphi|\psi(\varphi)|^{2\alpha}\right)^{1/\alpha}
\leq(2\pi)^{1/4\alpha-1/4\beta}\left(\sum_{m=-\infty}^{\infty}
|c_{m}|^{2\beta}\right)^{1/\beta}.\label{Ad20}
\end{align}
The coefficients $\alpha$ and $\beta$ fulfill the standard relations (\ref{mink}). Upon comparing the left hand side in (\ref{Ad20}) with the definition (\ref{angbin}), we see that we can use again the Jensen inequality for convex functions to obtain:
\begin{align} \delta\varphi^{1-\alpha}q_{n}^{\alpha}
=\delta\varphi^{1-\alpha}\left(\int_{n\delta\varphi}^{(n+1)\delta\varphi}d\varphi
|\psi(\varphi)|^{2}\right)^{\alpha}\leq\int_{n\delta\varphi}^{(n+1)\delta\varphi}d\varphi
|\psi(\varphi)|^{2\alpha}.\label{Ad21}
\end{align}
Next, we do summation over $n$:
\begin{align}
\delta\varphi^{1-\alpha}\sum_{n=0}^{N-1}q_{n}^{\alpha}
\leq\int_{0}^{2\pi}d\varphi|\psi(\varphi)|^{2\alpha}.\label{Ad22}
\end{align}
Putting the inequalities (\ref{Ad20}) and (\ref{Ad22}) together, we find:
\begin{align} (\delta\varphi)^{1/\alpha-1}\left(\sum_{n=0}^{N-1}q_{n}^{\alpha}\right)^{1/\alpha}
\leq(2\pi)^{1/4\alpha-1/4\beta}\left(\sum_{m=-\infty}^{\infty}|c_{m}|^{2\beta}\right)^{1/\beta}.
\label{Ad23}
\end{align}
This inequality will lead us directly to the final form of the uncertainty relation for the R\'enyi entropy. To this end let us define the R\'enyi entropy for angle and angular momentum in a following form:
\begin{align}
H_{\alpha}^{(\varphi)}=\frac{1}{1-\alpha}\ln\left[\sum_{n=0}^{N-1}q_{n}^{\alpha}\right],\label{Ad24}
\end{align}
\begin{align}
H_{\beta}^{(M)}=\frac{1}{1-\beta}\ln\left[\sum_{m=-\infty}^{\infty}p_{m}^{\beta}\right],\label{Ad25}
\end{align}
where $p_m=|c_m|^2$. Taking the logarithm of both sides of (\ref{Ad23}), multiplying by $\beta/(1-\beta)$, and identifying the R\'enyi entropies (\ref{Ad24}) and (\ref{Ad25}), we obtain final inequality:
\begin{align}
H_{\alpha}^{(\varphi)}+H_{\beta}^{(M)}\geq-\ln\frac{\delta\varphi}{2\pi}=\ln N.\label{Ad26}
\end{align}
As in the case of Maassen-Uffink result (\ref{Q69}), the right hand side of the inequality (\ref{Ad26}) is independent on the parameters $\alpha$ and $\beta$.

This uncertainty relation can be also applied to a different physical situation. Let us consider the state vectors of a harmonic oscillator expanded in the Fock basis,
\begin{align}
|\psi\rangle=\sum_{n=0}^{\infty}c_{n}|n\rangle.\label{Ad19a}
\end{align}
Every such state can be described by a function of $\varphi$ defined by \cite{ipb}:
\begin{align}
\psi(\varphi)=\sum_{n=0}^{\infty}c_{n}e^{im\varphi}.\label{Ad19b}
\end{align}
The only difference between this wave function and the wave function for a particle on a circle is the restriction to nonnegative values of $n$. Therefore, the inequalities (\ref{euram1}) and (\ref{Ad26}) still hold but they have now a different physical interpretation. The value of $n$ determines the amplitude and the angle $\varphi$ is the phase of the harmonic oscillation. This interpretation is useful in quantum electrodynamics where $n$ becomes the number of photons in a given mode and $\varphi$ is the phase of the electromagnetic field.

\subsubsection{The uncertainty relation for a particle on the surface of a sphere}

The uncertainty relation (\ref{Ad26}) may be called ``the uncertainty relation on a circle'', because the variable $\varphi$ varies from 0 to $2\pi$. We may ask, whether there is an uncertainty relation for a particle on the surface of a sphere. The answer is positive \cite{ibbmad} and to find the uncertainty relation we shall parameterize this surface by the azimuthal angle $\varphi\in[0,2\pi]$ and the polar angle $\theta\in[0,\pi]$. The wave functions on a sphere $\psi(\theta,\varphi)$ can be expanded into spherical harmonics:
\begin{align}
\psi(\theta,\varphi)
=\frac{1}{\sqrt{2\pi}}\sum_{l=0}^{\infty}\sum_{m=-l}^{l}c_{lm}Y_l^m(\theta,\varphi).\label{Ad27}
\end{align}
This expansion gives us a set of probabilities $|c_{lm}|^2$ which can be used to construct the R\'enyi entropy for the square of the angular momentum, determined by $l$, and for the projection of the angular momentum on the $z$ axis, determined by $m$. As in the case of a particle moving on a circle the two characteristics of the state --- angular position and angular momentum --- are complementary.

There is one mathematical property of spherical harmonics that will help us to derive the uncertainty relation. Namely, when $m=l$ then for large $l$ the function $Y_l^m$ is to a very good approximation localized in the neighborhood of the equator where $\theta=\pi/2$. In other words, if we divide the $\theta$ range into equal parts with length $\delta\theta$ then for every $\epsilon$ we can find sufficiently large $l$ such that the following relation holds (the modulus of the spherical harmonic does not depend on $\varphi$):
\begin{align}
\int_{\pi/2-\delta\theta/2}^{\pi/2+\delta\theta/2}\!\!d\theta|Y_l^l(\theta,\varphi)|^2=1-\epsilon.
\end{align}
This property follows from the fact that $|Y_l^l(\theta,\varphi)|^2=N|\sin\theta|^{2l}$. Of course, the smaller is the value of $\delta\theta$, the larger $l$ is required to localize the wave function.

To define the R\'enyi entropies in the standard way we need the probability distributions in angular momentum $p_{lm}=|c_{lm}|^{2}$ and in the position on the sphere
\begin{align} q_{ij}=\int_{i\delta\theta}^{(i+1)\delta\theta}\!\!\!d\theta\sin\theta
\int_{j\delta\varphi}^{(j+1)\delta\varphi}d\varphi\;|\psi(\theta,\varphi)|^{2}.\label{Ad29}
\end{align}
>From the argument about localizability of the wave function near the equatorial plane we deduce that for a {\em fixed} (sufficiently large) value of $l$ we can have the localization in one bin in the variable $\theta$. Therefore, $l$ and $\theta$ can be eliminated and we are left with the uncertainty relation in the projection of the angular momentum on the $z$ axis and the azimuthal angle $\varphi$. The uncertainty relation in these variables has already been established and it will have the same form (\ref{Ad26}) for the sphere. A complete proof of this fact is a bit more subtle and can be found in \cite{ibbmad}. The introduction of a preferred direction associated with the projection of the angular momentum is rather artificial and it would be interesting to find a rotationally invariant form of the uncertainty relation. It would also be of interest to find generalizations of the uncertainty relation (\ref{Ad26}) to more dimensions and to functions defined on manifolds different from the surface of a sphere.

\subsection{Uncertainty relations in phase space}

The next topic in this review is a ``phase-space approach'' to the uncertainty relations. We shall rewrite the sum of the Shannon entropies for position and momentum in a more symmetric way and we shall extend this symmetric description to the mixed states.

To this end we rewrite the sum of the expressions (\ref{csx}) and (\ref{csp}) for the Shannon entropies in the following, compact form:
\begin{align}
H^{(x)}+H^{(k)}
=-\sum_{i=-\infty}^{\infty}\sum_{j=-\infty}^{\infty}f_{ij}\ln\left(f_{ij}\right),\label{Ad10}
\end{align}
where:
\begin{align}
f_{ij}=q_ip_{j}=\int_{\left(i-1/2\right)\delta x}^{\left(i+1/2\right)\delta x}dx\:\int_{\left(j-1/2\right)\delta k}^{\left(j+1/2\right)\delta k}dk\:\left|\psi(x)\tilde{\psi}(k)\right|^{2}.\label{Ad11}
\end{align}
In this way, the uncertainty relation (\ref{eur}) becomes an inequality involving just one function $f_{1}(x,k)=\psi(x)\tilde{\psi}(k)^*$ defined on the phase space. This rearrangement is not so trivial when we extend it to the mixed states. For a density operator $\hat{\rho}$ we define the function:
\begin{align}
f(x,k)=\langle x|\hat{\rho}|k\rangle.\label{Ad12}
\end{align}
If the density operator represents a pure state, so that $\hat{\rho}=\left|\psi\rangle \langle \psi\right|$ we obtain the previous case $f(x,k)=f_{1}(x,k)$ but for mixed states the function (\ref{Ad12}) is not a simple product. Next, we define the two dimensional Fourier transform of $f(x,k)$:
\begin{align}
\tilde{f}(\lambda,\mu)&=\frac{1}{2\pi}\int_{\mathbb{R}}dx\,e^{-i\lambda x}\int_{\mathbb{R}}dk\,e^{ik\mu}f(x,k)\nonumber\\
&=\langle \lambda|\left(\int_{\mathbb{R}}dx\,|x\rangle\langle x|\right)\hat{\rho}\left(\int_{\mathbb{R}}dk\,|k\rangle\langle k|\right)|\mu\rangle .\label{Ad13}\nonumber\\
&=\langle\lambda|\hat{\rho}|\mu\rangle =\langle \mu|\hat{\rho}|\lambda\rangle^{*}=f(\mu,\lambda)^{*},
\end{align}
where we used the following representation of the Fourier transform kernels:
\begin{subequations}\label{Ad14}
\begin{align}
\frac{1}{\sqrt{2\pi}}e^{-i\lambda x}=\langle \lambda|x\rangle,\\
\frac{1}{\sqrt{2\pi}}e^{ik\mu}=\langle k|\mu\rangle.
\end{align}
\end{subequations}
The quantities in parentheses in this formula are the resolutions of the identity and were replaced by $1$. Thus, the function $f$ and its Fourier transform $\tilde{f}$ differ only in phase,
\begin{align}
\left|f\right|=\left|\tilde{f}\right|,\label{Ad16}
\end{align}
and we obtain:
\begin{align}
H^{(x,k)}=H^{\left(\mu,\lambda\right)}.\label{Ad17}
\end{align}
For the function $f_{1}$ this is, of course, a trivial conclusion. Because of (\ref{Ad17}) the uncertainty relation for the Shannon entropies becomes a relation that involves only one entropy, defined in terms of a single function $f$. The price paid for this simpler form of the uncertainty relation is the additional condition (\ref{Ad16}) that must be obeyed by the function $f(x,k)$.

\subsection{Mutually unbiased bases}

In (\ref{mut}) we have introduced the notion of mutually unbiased bases (MUB) and we discussed the Maassen-Uffink lower bound for these bases. In this subsection we are going to present several other uncertainty relations of this type. To this end, we introduce in a $D$-dimensional Hilbert space $\mathcal{H}_{D}$ a system of $M$  orthonormal bases $B^{m}=\left\{|b_i^{m}\rangle ,i\in 1,\ldots,D\right\} $, $m\in1,\ldots,M$. Two bases $B^{m}$ and $B^{n}$ are mutually unbiased bases if and only if \cite{survey} for each pair of vectors:
\begin{align}
\left|\langle b_i^{m}|b_{j}^{n}\rangle\right|^2 =\frac{1}{D}.\label{Ad36}
\end{align}
One may ask how many mutually unbiased bases can be found in $\mathcal{H}_{D}$? In other words how does the number $M_{\rm max}$ depends on $D$. There is no general answer to this question. It was found in \cite{ivonovic} that if $D$ is a prime number then $M_{\rm max}=D+1$.

We have already given the uncertainty relations (\ref{Q69}) for two MUB's related by the discrete Fourier transformation. A straightforward generalization of these relations involves a sum over several MUB's. Let us denote by $H^{(m)}$ the Shannon entropy for the $m$-th base:
\begin{align}
H^{(m)}=-\sum_{i=1}^{D}p_i^{m}\ln p_i^{m},\label{Ad37}
\end{align}
\begin{align}
p_i^{m}=\left|\langle b_i^{m}|\psi\rangle \right|^2.\label{Ad38}
\end{align}
Adding the uncertainty relations (\ref{Q69}) for every pair of MUB's, we obtain \cite{azar}:
\begin{align}
\sum_{m=1}^{M}H^{(m)}\geq\frac{M}{2}\ln D.\label{Ad39}
\end{align}
When $D$ is a prime number and we put $M=D+1$ then the relation (\ref{Ad39}) reads:
\begin{align}
\sum_{m=1}^{D+1}H^{(m)}\geq\frac{(D+1)}{2}\ln D,\label{Ad40}
\end{align}
but this bound is rather poor. S\'anchez \cite{san} found a better lower bound,
\begin{align}
\sum_{m=1}^{D+1}H^{(m)}\geq(D+1)\ln\frac{D+1}{2}.\label{Ad41}
\end{align}
The reader can find extensions of the uncertainty relations of this type in \cite{azar} and \cite{survey}. Recently in \cite{molmer} another refinement of the inequality (\ref{Ad39}) has been obtained,
\begin{align}
\sum_{m=1}^{M}H^{(m)}\geq M\,\ln\frac{MD}{M+D-1}.\label{Ad42}
\end{align}
For $M\geq1+\sqrt{D}$ this inequality is stronger than (\ref{Ad39}).

\subsection{Logarithmic Sobolev inequality}

In this subsection we find a direct relation between the standard Heisenberg uncertainty relation and the entropic uncertainty relation. This will be done with the use of a different mathematical tool --- the logarithmic Sobolev inequality and its inverse. We start with the inequalities for the continuous entropy (\ref{shx2}). The logarithmic Sobolev inequality for the Shannon entropy proved in \cite{sobol} reads:
\begin{align} S^{(x)}\geq\frac{1}{2}\left(1+\ln2\pi\right)-\frac{1}{2}\ln\left[L^{2}\int_{\mathbb{R}}dx
\frac{1}{\rho(x)}\left|\frac{d\rho(x)}{dx}\right|^{2}\right].\label{Ad31}
\end{align}
The important feature of this inequality is that we have only one function $\rho(x)$. Therefore, we do not need conjugate variables (momentum) to obtain a lower bound of the Shannon entropy (\ref{shx2}). On the other hand, the right hand side of (\ref{Ad31}) depends on $\rho(x)$, so it is a functional relation rather than an uncertainty relation. In order to obtain an uncertainty relation, we shall exploit the inverse logarithmic Sobolev inequality proved in \cite{invsob}:
\begin{align}
S^{(x)}\leq\frac{1}{2}(1+\ln2\pi)+\ln L\sigma_{x},\label{Ad32}
\end{align}
where $\sigma_{x}$ is the standard deviation (\ref{dispx}). Since this inequality involves only one distribution function, we can write it also for the probability density in momentum space ${\tilde\rho}(k)$:
\begin{align}
S^{(k)}\leq\frac{1}{2}(1+\ln2\pi)+\ln(\sigma_{k}/L),\label{Ad33}
\end{align}
Evaluating the exponential function of the sum of the inequalities (\ref{Ad32}) and (\ref{Ad33}), we obtain a refined version of the Heisenberg uncertainty relation \cite{bbm,rus}:
\begin{align}
\sigma_{x}\sigma_{k}
\geq\frac{1}{2}\exp\left(S^{(x)}+S^{(k)}-1-\ln\pi\right)\geq\frac{1}{2}.\label{Ad35}
\end{align}
This uncertainty relation is stronger than the standard Heisenberg uncertainty relation. Whenever the sum of the Shannon entropies exceeds its lower bound $1+\ln\pi$, we obtain a stronger bound for $\sigma_{x}\sigma_{k}$ than the standard $1/2$. Note that the uncertainty relation (\ref{Ad35}) holds for any pair of distributions. They do not have to be related by the Fourier transformation of the wave functions.

\end{document}